\documentclass[prl, superscriptaddress, twocolumn, showpacs]{revtex4}

\usepackage{amsmath, amssymb, braket}
\usepackage{mathrsfs}
\usepackage{epstopdf}
\usepackage{graphicx, pdflscape, array, makecell}
\usepackage[sort&compress]{natbib}
\usepackage{natmove}
\usepackage[pdftex = true, colorlinks = true, citecolor = blue, bookmarks = true]{hyperref}
\usepackage{hypernat}
\usepackage[utf8]{inputenc}
\usepackage[OT4]{fontenc}

\newcommand{\beq}{\begin{equation}}

\newcommand{\eeq}{\end{equation}}

\newcommand{\bec}{\begin{center}}

\newcommand{\eec}{\end{center}}

\begin{document}


\title{A DFT Approach to Non-Covalent Interactions via Monomer Polarization and Pauli Blockade}
\date{\today}
\author{Łukasz Rajchel}
\email{lrajchel@tiger.chem.uw.edu.pl}
\affiliation{Department of Chemistry, Oakland University, Rochester, Michigan 48309-4477, USA}
\affiliation{Faculty of Chemistry, University of Warsaw, 02-093 Warszawa, Pasteura 1, Poland}
\author{Piotr S. Żuchowski}
\affiliation{Department of Chemistry, Durham University, South Road, Durham DH1 3LE, United Kingdom}
\author{Małgorzata M. Szczęśniak}
\affiliation{Department of Chemistry, Oakland University, Rochester, Michigan 48309-4477, USA}
\author{Grzegorz Chałasiński}
\email{chalbie@tiger.chem.uw.edu.pl}
\affiliation{Department of Chemistry, Oakland University, Rochester, Michigan 48309-4477, USA}
\affiliation{Faculty of Chemistry, University of Warsaw, 02-093 Warszawa, Pasteura 1, Poland}


\begin{abstract}
We propose a~"DFT+dispersion" treatment which avoids double counting of dispersion terms by deriving the dispersion-free density functional theory~(DFT) interaction energy and combining it with DFT-based dispersion. The formalism involves self-consistent polarization of DFT monomers restrained by the~exclusion principle via the~Pauli blockade technique. Any exchange-correlation potential can be used within monomers, but only the exchange operates between them. The applications to rare-gas dimers, ion-rare gas interactions and hydrogen bonds demonstrate that the interaction energies agree with benchmark values.
\end{abstract}

\pacs{31.15.E-, 34.20.Gj}

\maketitle


The applicability of the density functional theory~(DFT) to calculations of intermolecular potentials of van~der~Waals complexes depends upon a seamless inclusion of the dispersion energy, a long-range correlation effect, in the~DFT treatment. This goal has been pursued vigorously along many lines~\cite{lund_ander_shao_chan_lang:1995, grimme:2004, xu_goddard:2004, lilien_tavern_roth_sebas:2004, truhlar_zhao:2006, toul_gerber_jansen_savin_angyan:2009, janesko_hend_scus:2009} with varying success. One promising avenue consists of using an \textit{a posteriori} dispersion correction added to supermolecular DFT calculations of interaction energy, in the spirit of~\citet{ahlr_penco_scoles:1977}. For this strategy to succeed two elements are necessary: an accurate, nonempirical description of dispersion energy between DFT monomers and a sensible dispersion-free description of supermolecular interaction energy within DFT.

A highly accurate and computationally efficient formulation of dispersion energy is now available from the time-dependent DFT as proposed by~\citet{mis_jez_szal:2003} and \citet{hess_jansen:2003}, referred to as coupled Kohn-Sham~(CKS) dispersion. The second element of this strategy, however, has not been available up to now. In contrast to the Hartree-Fock~(HF) interaction energy which is well defined and contains physically interpretable effects, the analogous DFT interaction energy has neither of these characteristics. Depending on a particular functional it may include a variety of obscure terms. In particular, for all current generalized gradient approximation~(GGA) formalisms no dispersion contribution is accounted for at large intemolecular separations~$R$ and only some part appears at intermediate and small~$R$. As a consequence, approximate exchange-correlation (xc) functionals often exhibit an artifactual behavior in the long range as well as in the van~der~Waals minimum region. A rigorous DFT+dispersion approach should be based on a DFT interaction energy that \textit{a priori} avoids these residual dispersion terms but allows for accurate mutual exchange and polarization effects --- an analog of the~HF interaction energy at the DFT level of theory. Such a DFT interaction energy could then be combined with a dispersion component obtained from CKS or other formalisms~\cite{becke_johnson:2007, dion_rydb_schr_langr_lund:2004}.

To this end we adapt the formalism of Pauli-blockade Hartree-Fock~(PB~HF)~\cite{gut_pauli:1988} combined with the bifunctional subsystem formulation of DFT of~\citet{rajchel_zuch_szczesniak_chal:2010_cpl} Specifically, the energy of the complex is evaluated from the classic Heitler-London formula which takes the antisymmetrized product of participating monomer wavefunctions while the monomers are described with the KS orbitals from DFT calculations. In the second step, one iteratively evaluates the interaction energy between two DFT monomers, described by KS determinants, in a manner analogous to the HF method. That is, the monomers are polarized in each other's fields until self-consistency under the constraint of the Pauli exclusion principle between monomers. Within the monomers any exchange-correlation DFT potential may be employed, whereas between monomers the exact exchange potential is used to avoid the dispersion contribution. In the third step, the dispersion component is \textit{a posteriori} added. In such a way the erratic behavior of approximate xc functionals is eliminated and all physically important effects are included without the problem of the dispersion double counting. Note that the concept of the separated monomers resembles the idea of the "range-separation" approach in DFT methodology~\cite{angyan_gerber_savin_toul:2005}.

In the present Letter we present an outline of the application of the PB scheme to the calculation of the interaction energy of DFT monomers. More details on derivation and implementation will be published elsewhere. The procedure starts with KS solutions for the isolated monomers~A and~B. The solution is obtained in a self-consistent way involving coupled equations for both subsystems. The equation for monomer~A reads:
	\beq
		\left( \hat{\tilde{f}}_\mathrm{A}^{\mathrm{KS}[n]} +
			\hat{\tilde{v}}_\mathrm{B}^{\mathrm{elst}{[n]}} + \hat{\tilde{v}}_\mathrm{B}^{\mathrm{exch}{[n]}} +
			\eta \hat{\tilde{R}}_\mathrm{B}^{[n]} \right) a_i^{[n + 1]}
			= \epsilon_{\mathrm{A}, i}^{[n + 1]} a_i^{[n + 1]} \\
		\label{eq:PBiter}
	\eeq
and its monomer~B counterpart is simply generated through the exchange of~A and~B indices. In~Eq.~\eqref{eq:PBiter}, $[n]$ superscripts denote iteration numbers,
	\beq
		\hat{\tilde{f}}_\mathrm{A}^\mathrm{KS[n]}(\mathbf{r}) = \hat{h}_\mathrm{A}(\mathbf{r}) + \int_{\mathbb{R}^3} \frac{\tilde{\rho}_\mathrm{A}^{[n]} (\mathbf{r}')}{|\mathbf{r} - \mathbf{r}'|} \, d^3 \mathbf{r}' +
			\tilde{v}_\mathrm{A}^\mathrm{xc[n]}(\mathbf{r}),
		\label{eq:KSop}
	\eeq
is the standard KS operator~(e.g., see~Ref.~\onlinecite{koch_holt:2001}) built from the~monomer~A orbitals~$\left( a_i^{[n + 1]} \right)$, the two following terms are electrostatic and exchange HF potentials, and the last term is the so-called penalty operator enforcing the proper mutual orthogonality between~A and~B~monomer orbitals so that the intersystem Pauli exclusion principle is fulfilled. $\eta > 0$ is a parameter that does not affect the final solution. Monomer~A density in terms of KS orbitals is
	\beq
		\rho_\mathrm{A}^{[n]}(\mathbf{r}) = 2 \sum_{i \in A} \left| a_i^{[n]}(\mathbf{r}) \right|^2.
	\eeq
The xc potential in~\eqref{eq:KSop} requires the asymptotic correction of~\citet{grun_grits_gisb_baer:2001} in order to yield appropriate interaction energies. The orbitals are symmetrically orthogonalized after each iteration, yielding~$\left\{ \tilde{a}_i^{[n + 1]} \right\}_{i \in A}, \left\{ \tilde{b}_k^{[n + 1]} \right\}_{k \in B}$ set. The tilde sign~(e.g.~$\tilde{E}_\mathrm{int}^\mathrm{PB}$) denotes quantities calculated with such orbitals. The interaction energy at each iteration is decomposed into several terms:
	\beq
		E_\mathrm{int}^{\mathrm{PB}[n]} = \Delta \tilde{E}_\mathrm{A}^{[n]} + \Delta \tilde{E}_\mathrm{B}^{[n]} + \tilde{E}_\mathrm{elst}^{[n]} + \tilde{E}_\mathrm{exch}^{[n]}.
		\label{eq:PB_enint}
	\eeq
$\Delta \tilde{E}_\mathrm{X}^{[n]} = E_\mathrm{X} \left[ \tilde{\rho}_\mathrm{X}^{[n]} \right] - E_\mathrm{X} \left[ \rho_\mathrm{X}^0 \right]$ terms are a result of the imposition of the intersystem Pauli exclusion principle and contain repulsion energy. The expressions for electrostatic and exchange energies are very simple due to the orthogonality of orbitals:
	\beq
		\begin{split}
			\tilde{E}_\mathrm{elst}^{[n]} & =
				\int_{\mathbb{R}^3} v_\mathrm{A}^\mathrm{ne}(\mathbf{r}) \tilde{\rho}_\mathrm{B}^{[n]}(\mathbf{r}) \, d^3 \mathbf{r} +
				\int_{\mathbb{R}^3} v_\mathrm{B}^\mathrm{ne}(\mathbf{r}) \tilde{\rho}_\mathrm{A}^{[n]}(\mathbf{r}) \, d^3 \mathbf{r} + \\
				& + 4 \sum_{i \in A} \sum_{k \in B} \Braket{ \tilde{a}_i^{[n]} \tilde{b}_k^{[n]} | \tilde{a}_i^{[n]} \tilde{b}_k^{[n]} } + V_{\mathrm{AB}}^\mathrm{nn},
		\end{split}
	\eeq
and
	\beq
		\tilde{E}_\mathrm{exch}^{[n]} = -2 \sum_{i \in A} \sum_{k \in B} \Braket{ \tilde{a}_i^{[n]} \tilde{b}_k^{[n]} | \tilde{b}_k^{[n]} \tilde{a}_i^{[n]} },
	\eeq
where $v_\mathrm{A}^\mathrm{ne}$ is the potential due to monomer~A nuclei and~$V_{\mathrm{AB}}^\mathrm{nn}$ the intermonomer nuclear-nuclear repulsion term. The zero-iteration energy~\eqref{eq:PB_enint} is the DFT analog of the well-known Heitler-London interaction energy:~$E_\mathrm{int}^{\mathrm{PB}[0]} \equiv \mathscr{E}_\mathrm{int}^\mathrm{HL}$ (cf. also ~\citet{cyb_seversen:2005}). The HL energy contains the well defined electrostatic and exchange interaction contributions, in this case between unperturbed isolated DFT monomers. It is closely related to the first order energy in the symmetry-based perturbation theory based on DFT~SAPT(DFT). Coupling between the subsystems via~Eq.~\eqref{eq:PBiter} leads to the system energy lowering, and is referred to as deformation energy:
	\beq
		\mathscr{E}_\mathrm{def}^\mathrm{PB} = \mathscr{E}_\mathrm{int}^\mathrm{PB} - \mathscr{E}_\mathrm{int}^\mathrm{HL}.
	\eeq
$\mathscr{E}_\mathrm{int}^\mathrm{PB}$ represents the final PB energy, calculated with~\eqref{eq:PB_enint} using self-consistent orbitals satisfying~\eqref{eq:PBiter}. It results from the mutual electric polarization of DFT monomers, and, owing to the Pauli blockade procedure, it contains exchange contributions. It is related to the induction terms of the SAPT formalism with two important advantages over the latter: PB gathers all electric polarization terms to infinity and accounts for accompanying exchange effects in a consistent manner within the DFT formalism.

\begin{figure}[htbp]
	\bec
	\includegraphics[width = 8.6cm]{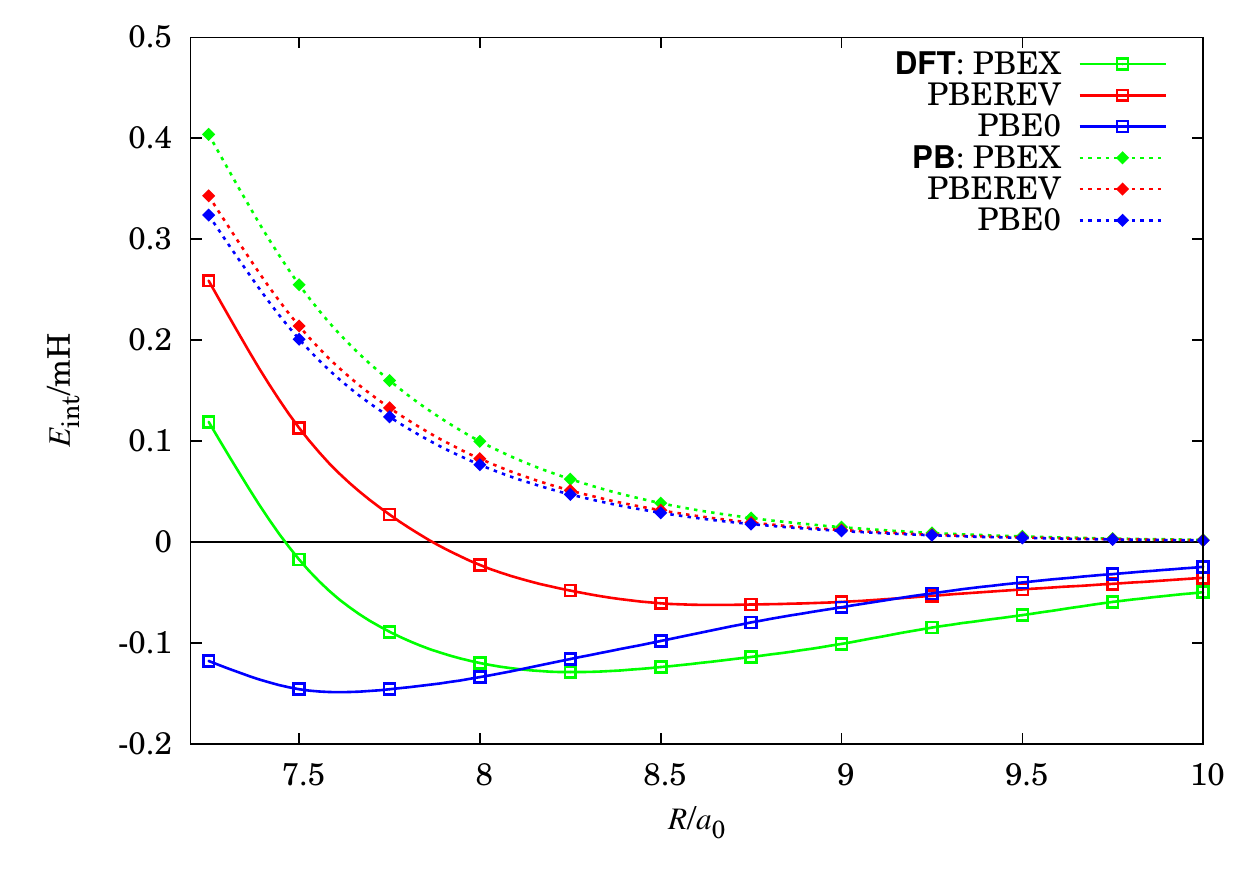}
	\caption{(Color online) Comparison of PB and supermolecular DFT interaction energies for~Ar$_2$.}
	\label{fig:Ar2_PBDFT}
	\eec
	\end{figure}

In the original HF-based formulation, the~PB procedure simply restores the supermolecular HF interaction energy, and obviously neglects any kind of electron correlation. For the DFT analog, both monomers are described with the full KS operator, but are coupled using HF coulomb and exchange operators~[$\hat{\tilde{v}}_\mathrm{B}^{\mathrm{elst}}$ and~$\hat{\tilde{v}}_\mathrm{B}^{\mathrm{exch}}$ in~Eq.~\eqref{eq:PBiter}, respectively] built from KS orbitals. Such an approach accounts for intramonomer local electron correlation, leaving out the intermonomer nonlocal contributions. The~$\mathscr{E}_\mathrm{int}^\mathrm{PB}$ represents then "non-dispersion" part of the interaction energy that includes the electrostatic, exchange, and induction components. For rare gas dimers it is purely repulsive. In Fig.~\ref{fig:Ar2_PBDFT} $\mathscr{E}_\mathrm{int}^\mathrm{PB}$ is compared with the supermolecular counterpoise corrected DFT interaction energy for Ar$_2$. 
Calculations employed three DFT functionals of the GGA PBE hierarchy, which systematically improve on the description of monomers: PBEX~\cite{perdew_burke_ernz:1996}~(exchange only), PBEREV~\cite{zhang_yang:1998} (local exchange plus correlation), and PBE0~\cite{adamo_barone:1999} (a hybrid of local and exact exchange plus correlation). All calculations used aug-cc-pVTZ basis sets. As seen in Fig.~\ref{fig:Ar2_PBDFT} $\mathscr{E}_\mathrm{int}^\mathrm{PB}$ is indeed purely repulsive for Ar$_2$. By contrast, the supermolecular DFT interaction energies reveal minima of an~unknown origin.

The total interaction energy, termed PBD for Pauli blockade plus dispersion, is obtained by adding to~$\mathscr{E}_\mathrm{int}^\mathrm{PB}$ the dispersion component obtained from CKS or other accurate techniques~\cite{becke_johnson:2007, dion_rydb_schr_langr_lund:2004}:
	\beq
		E_\mathrm{int}^\mathrm{PBD} = \mathscr{E}_\mathrm{int}^\mathrm{PB} + E_\mathrm{disp}.
		\label{eq:PBD_int}
	\eeq
In this work we use the second-order CKS dispersion components combined with the exchange-dispersion term from SAPT(DFT)~\cite{mis_szal:2002, mis_jez_szal:2003, jans_hess:2001}.

$E_\mathrm{int}^\mathrm{PBD}$ can now be compared with SAPT(DFT) theory. SAPT(DFT) interaction energy
	\beq
		E_\mathrm{int}^{\mathrm{SAPT}\delta} = E^{(1)} + E^{(2)} + \delta_\mathrm{HF}
		\label{eq:SAPTd_int}
	\eeq
includes first-order (electrostatic and exchange) and second-order (induction, dispersion and their exchange counterparts) contributions derived from KS orbitals. $\delta_\mathrm{HF}$ denotes approximate correction for higher-order induction effects derived at the~HF level of theory~\cite{mis_pod_jez_szal:2005}. The SAPT(DFT) terms were calculated using MOLPRO program~\cite{MOLPRO_brief_2009}.

\begin{figure}[htbp]
	\bec
	\includegraphics[width = 8.6cm]{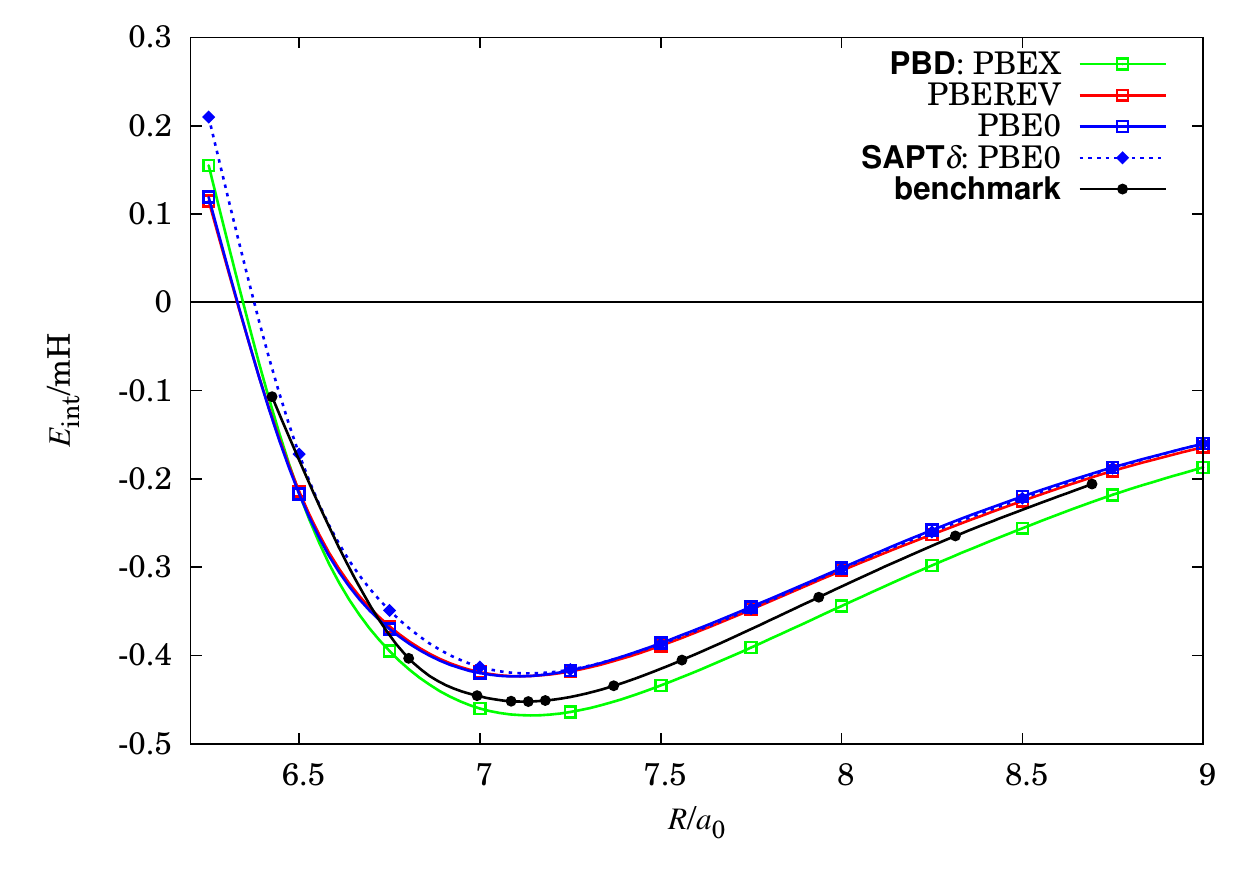}
	\caption{(Color online) Comparison of PBD and SAPT$\delta$ interaction energies for~Ar$_2$.}
	\label{fig:Ar2_PBDSAPTd}
	\eec
	\end{figure}

\begin{figure}[htbp]
	\bec
	\includegraphics[width = 8.6cm]{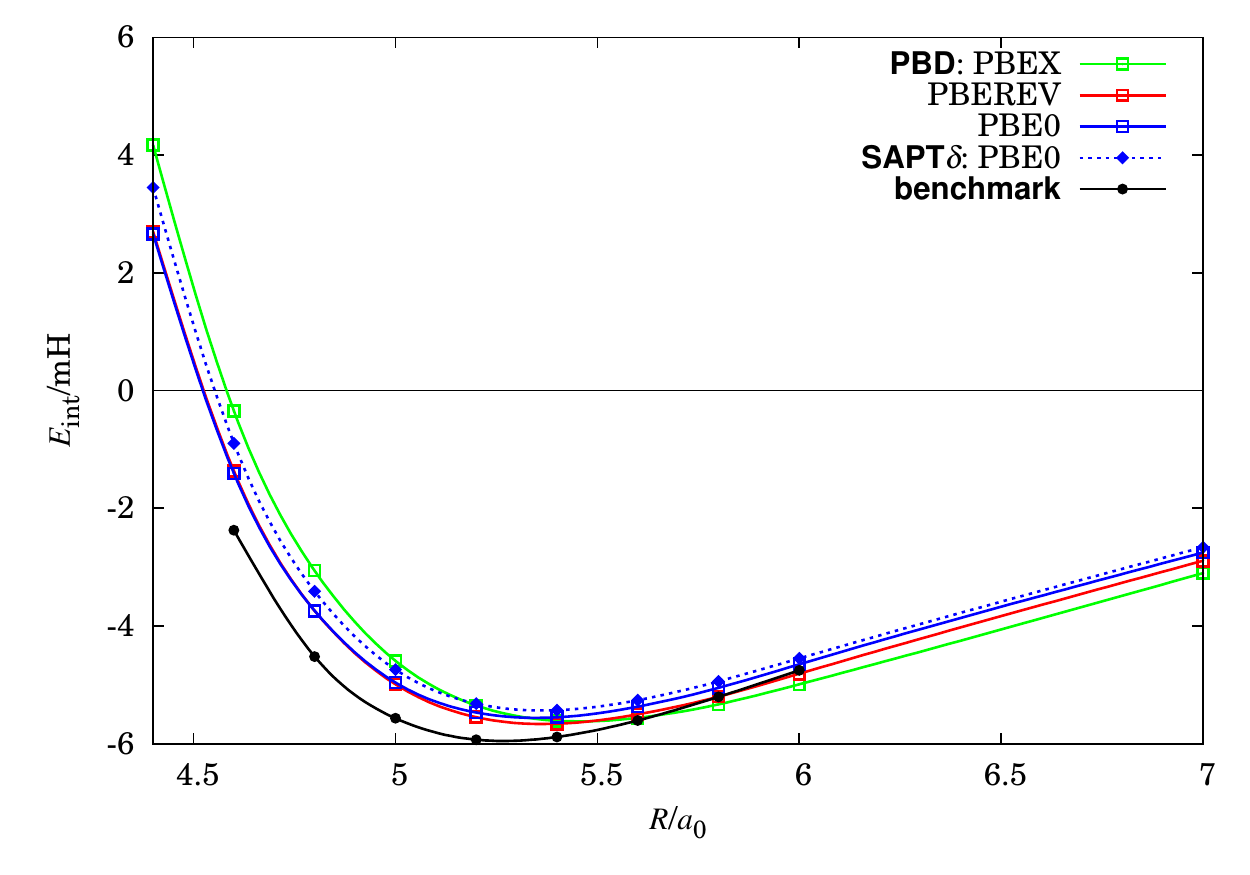}
	\caption{(Color online) Comparison of PBD and SAPT$\delta$ interaction energies for~ArNa$^+$.}
	\label{fig:ArNa+_PBDSAPTd}
	\eec
	\end{figure}

\begin{figure}[htbp]
	\bec
	\includegraphics[width = 8.6cm]{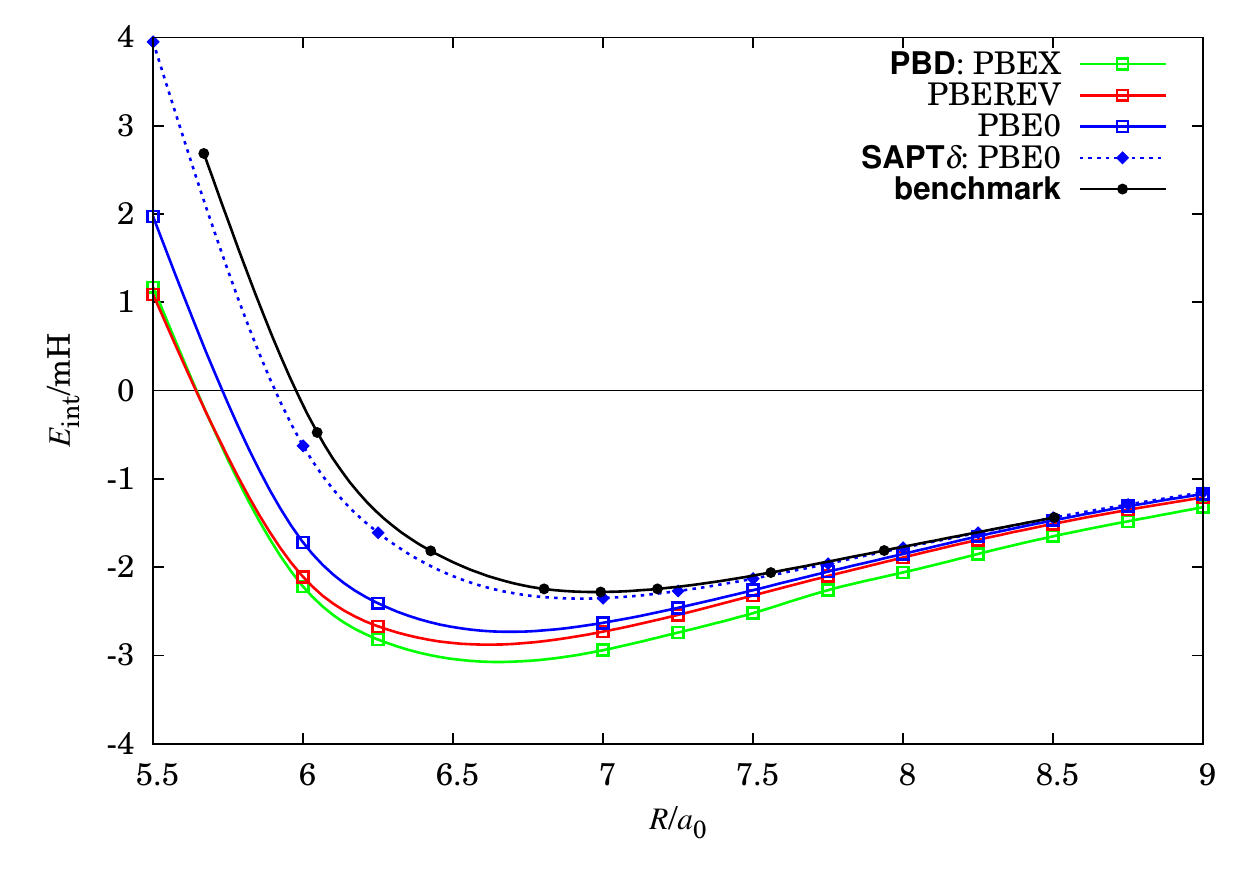}
	\caption{(Color online) Comparison of PBD and SAPT$\delta$ interaction energies for~ArCl$^-$.}
	\label{fig:ArCl-_PBDSAPTd}
	\eec
	\end{figure}

To test the efficiency of the PBD approach, we performed calculations for several diatomic systems composed of closed-shell atoms and ions: Ar$_2$, ArNa$^+$, and ArCl$^-$. Again, the same set of PBE functionals has been used with the same aug-cc-pVTZ basis set. The comparison of energies calculated with~Eqs.~\eqref{eq:PBD_int} and~\eqref{eq:SAPTd_int} is presented in~Figs.~\ref{fig:Ar2_PBDSAPTd}~through~\ref{fig:ArCl-_PBDSAPTd}.

The results are compared with high-level \textit{ab initio} data from Refs.~\onlinecite{patk_murd_fou_szal:2005}~(Ar$_2$), \onlinecite{rez_alm_roegg:1995}~(ArNa$^+$), and~\onlinecite{buch_krems_szcz_xiao_vieh_chal:2001}~(ArCl$^-$). One can see a remarkably good agreement of PBD and benchmark data, similar to that of SAPT$\delta$, for Ar$_2$, and ArNa$^+$. The agreement for ArCl$^-$ is somewhat worse, however for this system the benchmark calculation may be up to 10~\% too shallow.

Another test of the proposed approach was performed for hydrogen-bonded and other molecular complexes. The results for a representative set of these complexes are shown in~Table~\ref{tab:en_int_Hb}.
	\begin{table}[htbp]
\caption{Comparison of interaction energies and their components from PBD calculations with SAPT(DFT) and benchmark values~(in~mH). Equilibrium geometries are from indicated references. PBE0 functionals with aug-cc-pVTZ basis set are used for monomers (aug-cc-pVDZ for CO$_2$).}
\label{tab:en_int_Hb}
\bec
\begin{tabular}{l*{5}{>{$}c<{$}}>{$}r<{$}}
\hline \hline
System & \mathscr{E}_\mathrm{int}^\mathrm{HL}\footnote{Electrostatic + exchange.} & \mathscr{E}_\mathrm{int}^\mathrm{PB}\footnote{Electrostatic + exchange + mutual polarization.} & E_\mathrm{int}^\mathrm{PBD}\footnote{Electrostatic + exchange + mutual polarization + dispersion.} & \delta_\mathrm{HF}\footnote{Higher-order induction.} & E_\mathrm{int}^{\mathrm{SAPT}\delta} & $Benchmark$ \\
\hline
(H$_2$O)$_2$ & 	-1.02 & -4.94 & -8.56 & -1.45 & -8.12 
 & -9.00$~\cite{jur_cerny_hobza:2006}$ \\
(HF)$_2$ & 	-0.693 & -4.78 & -7.59 & -1.19 & -6.48 
 & -7.22$~\cite{halk_klop_helg_jorg_taylor:1999}$ \\
(HCl)$_2$ & 	1.65  & -0.548 & -3.48 & -0.976 & -3.16 
 & -3.10$~\cite{halk_klop_helg_jorg_taylor:1999}$ \\
(NH$_3$)$_2$ & 	-0.479 & -2.03 & -5.34 & -0.522 & -4.78 
 & -5.05$~\cite{jur_cerny_hobza:2006}$ \\
H$_2$O$-$HF & 	-1.15 & -10.2 & -14.8 & -3.02 & -13.2 
 & -13.55$~\cite{halk_klop_helg_jorg_taylor:1999}$ \\
(CH$_4$)$_2$ & 	0.705 & 0.631 & -0.852 & -0.0378 & -0.948 
 & -0.813$~\cite{jur_cerny_hobza:2006}$ \\
(CO$_2$)$_2$ & 	0.85  & 0.501 & -1.74 & -0.0673 & -2.11 
 & -1.80\footnote{CCSD(T) result for the global minimum from~\cite{buk_sadlej_jez_jank_szal_kuch_will_rice:1999}.}~$\cite{buk_sadlej_jez_jank_szal_kuch_will_rice:1999}$ \\
\hline \hline
\end{tabular}
\eec
\end{table}
The~PBD interaction energies are compared with SAPT$\delta$ and with benchmark values. PBD and SAPT$\delta$ share the same values of dispersion energy. The benchmarks correspond to basis set saturated CCSD(T) results. The comparison indicates that PBD leads to very reasonable interaction energies for hydrogen-bonded systems. As mentioned above, the two components $\mathscr{E}_\mathrm{int}^\mathrm{HL}$ and $\mathscr{E}_\mathrm{int}^\mathrm{PB}$ have a~clear physical interpretation of electrostatic-plus-exchange interaction of two unperturbed monomers and mutual monomer polarization contribution restrained by exchange, respectively. It should also be noted that whereas SAPT$\delta$ provides results of equally high quality, it is dependent upon inclusion of~$\delta_\mathrm{HF}$ which is substantial.

In summary, we presented a new treatment of interaction energy between two DFT monomers which is exactly dispersion-free. It contains two physically-meaningful terms: HL interaction energy and self-consistent polarization which is restrained by the exact exchange. This interaction energy which is an analog of the supermolecular HF interaction energy can be combined with existing, reliable treatments of dispersion to yield the first theoretically sound DFT+dispersion approach. The computational cost of our approach is essentially the same as that of the routine KS calculations, and it scales with the size and number of fragments as the standard KS equations. It is worthwhile to note that for the test cases of this Letter the results proved to be remarkably insensitive to the asymptotic correction --- an issue which deserves further studies. 

Financial support from NSF (Grant~No.~CHE-0719260) is gratefully acknowledged. We acknowledge computational resources purchased through NSF~MRI program (Grant~No.~CHE-0722689). This work was sponsored by the Polish Funds for Sciences (Grant~No.~N204/108335).


\bibliography{Paper-PRL-arXiv}

\end{document}